\documentclass[twocolumn,secnumarabic,amssymb,nobibnotes,nofootinbib,aps, prd]{revtex4-1}
\pdfoutput=1

\usepackage{graphicx}
\usepackage{enumitem}
\usepackage{latexsym}
\usepackage{amsfonts}
\usepackage{amssymb}
\usepackage{color}
\usepackage{amsmath}
\usepackage{slashed}
\usepackage{dcolumn}
\usepackage{verbatim}

\definecolor{darkblue}{rgb}{0,0,0.5}
\usepackage[
pdfauthor={Nirmal Raj}]{hyperref}

\DeclareGraphicsExtensions{.pdf,.png,.jpeg}

\newcommand{\beq}{\begin{equation}}
\newcommand{\eeq}{\end{equation}}
\newcommand{\bea}{\begin{eqnarray}}
\newcommand{\eea}{\end{eqnarray}}

\newcommand{\sg}{\widetilde{g}}
\newcommand{\sq}{\widetilde{q}}
\newcommand{\scolor}{\widetilde{C}}
\newcommand{\dm}{\widetilde{\chi}_1^0}
\newcommand{\met}{\slashed{E}_T}
\newcommand{\Etag}{E^{\rm ISR}_{\rm tag}}
\newcommand{\msg}{m_{\tilde{g}}}
\newcommand{\mdm}{m_{\tilde{\chi}^0}}
\newcommand{\delm}{\Delta M}
\def\newvar{\rho}
\def\meteight{\slashed{E}_{T,8}^{\rm cut}}
\def\metthirt{\slashed{E}_{T,13}^{\rm cut}}

\def\shat{\hat{s}}

\newcommand{\gsim}{\gtrsim}
\newcommand{\lsim}{\lesssim}
\newcommand{\ra}{\rightarrow}

\allowdisplaybreaks

\begin{document}	

\title{Extending the Reach of Compressed Gluinos at the LHC}

\author{Antonio Delgado}
\author{Adam Martin}
\author{Nirmal Raj}
\affiliation{Department of Physics, University of Notre Dame, 225 Nieuwland Hall, Notre Dame, Indiana 46556, USA}

\begin{abstract}
Conventional supersymmetry searches rely on large missing momentum and, on that account, are unsuitable for discovering superpartners nearly degenerate with the LSP. 
Such ``compressed regions" are best probed by dedicated strategies that exploit their unique kinematic features. 
We consider a case study of a compressed gluino-bino simplified spectrum, motivated by its ability to set the dark matter relic abundance via co-annihilation.
A kinematic variable suited to this spectrum is introduced, by which, for a gluino-bino mass splitting of 100 GeV, the discovery reach is extendable to $\msg$ = 850 GeV (1370 GeV) at LHC center-of-mass energy 8 TeV (13 TeV) with luminosity 20 fb$^{-1}$ (3000 fb$^{-1}$).
The non-trivial role played by soft triggers is also discussed.
\end{abstract}

\maketitle

\section{Introduction}
\label{sec:intro}

Supersymmetry remains the most popular resolution to the question of stabilizing the electroweak scale.
Furthermore, scenarios containing R-parity are highly motivated to avoid rapid proton decay; they give as subproduct a dark matter (DM) candidate in the form of the lightest supersymmetric particle (LSP).
Search signatures conventionally rely on large missing transverse momentum (MET) coming from the LSP accompanied by energetic visible states such as jets, leptons and photons.
This approach, however, fails for a superpartner spectrum of near-degenerate states, the so-called compressed regions of parameter space.
Due to poor sensitivities to subdued MET and visible energies, the kinematics of such spectra are not easily distinguished from Standard Model (SM) production.
Compressed spectra, at the same time, hold cosmological interest as they play a unique role in setting the abundance of dark matter.
They can possibly modify the effective thermal cross-section at the point of DM freezeout by the mechanism of co-annihilation \cite{Griest:1990kh}.
If weak scale supersymmetry were to hide in such parametric regions, it is an important challenge to better our strategies for enhancing search sensitivities.

In the case of colored superpartner production, collider searches for compressed spectra are plagued by another problem.
Consider a simplified model with a colored superpartner $\scolor$ and a neutralino LSP $\dm$ that are nearly mass degenerate. 
The jets that $\scolor$ decay into are generally soft, and thus the event is contaminated with initial state radiation (ISR), confounding jets+MET-based searches. 
However, this feature can be turned into an advantage.
Monojet searches are capable of looking for particles carrying missing energy recoiling against the ISR that dominates compressed regions 
\cite{Aad:2014nra,
Khachatryan:2015wza,
CMSstopcharm,
Aad:2015zva,
Low:2014cba}; 
the sensitivity, unfortunately, remains poor.
In the case of compressed stop searches, some headway has been made by availing spin correlations \cite{Aad:2014mfk}, an approach shown to outdo traditional searches when the stop is near-degenerate with the top quark in mass.
Recent papers 
\cite{Hagiwara:2013tva,
An:2015uwa,
Cheng:2016mcw,
Macaluso:2015wja} 
have suggested searching for generic compressed stop regions ($m_{\tilde{t}} \approx \mdm + m_t$) by taking advantage of the ISR-domination together with the fact that heavy stops and LSPs are slow-moving in the lab frame.

In this paper, we will adopt the latter approach to hunt for other colored superpartners; we will improve upon it, and show that search sensitivities for such partners residing in compressed spectra can be enhanced significantly, even using existing Run I data.
As a case study, we will pick a simplified model with a gluino taken slightly heavier than a bino LSP.
We make this choice motivated by its well-known viability in setting the relic abundance of DM by means of co-annihilation
\cite{Profumo:2004wk,
Feldman:2009zc,
deSimone:2014pda,
Harigaya:2014dwa,
Ellis:2015vaa,
Nagata:2015hha,
Ellis:2015vna}
and the relative safety of bino DM from limits set by direct detection experiments.
The central idea of gluino-bino co-annihilation is that while binos self-annihilate inefficiently, naively prompting overclosure of the universe,
the simultaneous annihilation of gluinos in the thermal bath can effectively help drive the relic abundance down to the observed value.  
Analogous co-annihilation scenarios involving squarks, including stop-LSP and sbottom-LSP co-annihilation, are also possible.
These can also be covered by our collider search strategy.
Our strategy can also be easily generalized to a new (non-supersymmetric) particle of any color structure and spin -- the possibilities are itemized in \cite{deSimone:2014pda} -- and to Dirac masses.
(For a comprehensive list of co-annihilation spectra in simplified DM models see \cite{Baker:2015qna}.) 
We leave such generalizations to future work. 

The calculation of the precise relic abundance $\Omega_\chi h^2$ of the gluino-bino spectrum must take into account the effects of the co-annihilation process, Sommerfeld enhancement and the formation of gluino-gluino bound states that may decouple from the plasma \cite{Ellis:2015vaa}.
These effects determine, for a given gluino mass, the size of the gluino-LSP mass splitting $\delm$, usually computed numerically.
In our analysis, we will not concern ourselves with the precise determination of $\delm$ required for the observed $\Omega_\chi h^2$, since our focus is on the collider search.
For gluinos lighter than 8 TeV, it has been computed that $\delm$ ranges between a few 10's of GeV and 140 GeV 
\cite{deSimone:2014pda,
Harigaya:2014dwa,
Ellis:2015vaa,
Nagata:2015hha}. 
Motivated by this, we set  
\beq
\delm = 100~{\rm GeV}
\label{eq:splitting}
\eeq
throughout the analysis. 

In our simplified model, squarks, charginos and the other neutralinos are decoupled from the spectrum and their masses set at 10 TeV.
We remark that taking squarks any heavier can lead to two consequences:

(1) As explained in \cite{Ellis:2015vaa} and \cite{Nagata:2015hha}, if the squarks are $\gtrsim 100$ times heavier than the gluino-bino system, the rate of interconversion between gluinos and binos at the time of freezeout becomes so slow that the number densities of either species become uncorrelated and evolve independent of each other.
The validity of the co-annihilation calculation is then compromised.

(2) $\mathcal{O}(100~{\rm TeV})$ squarks suppress the decay width of the gluino, such that for $\delm \sim 100$~GeV, the decay length  $\gsim \mathcal{O}$(mm). 
As demonstrated in \cite{Nagata:2015hha}, this region can be probed by displaced vertices at the LHC.
Since our focus is on prompt decays, we wish to avoid this region, although it must be noted that our strategy can complement displaced vertex searches if the squarks are very heavy.

With all superpartners besides the gluino and bino decoupled, gluino pair production will proceed dominantly via QCD, so the sole free parameter in our analysis is the gluino mass, $\msg$.
The simplified model as we have described here corresponds to the CMS model ``T1qqqq" with 
the decay topology $\sg \ra q q \dm$.
Current limits set by ATLAS and CMS on this topology at the 8 TeV LHC \cite{Aad:2014wea,Aad:2015iea,Chatrchyan:2013mys,CMS:2014wsa,CMS:2014ksa} have ruled out $\msg \lsim 600$~GeV for $\delm = 100$~GeV, and hence we will deal with $\msg \geq 600$~GeV.
As a comparison, we remind the reader that the bound on this scenario for a massless LSP is $\msg \gsim 1400$~GeV.

There has been some recent theory literature exploring the compressed gluino scenario. For instance,
Ref.~\cite{Han:2015lha} studies the use of ``fat jets" that capture the partons from gluino decay.  
Ref.~\cite{Chalons:2015vja} looks for the radiative decay $\sg \ra {\rm gluon} + \dm$ in special parametric regions. 
The applicability of variables already in experimental use is explored in \cite{Bhattacherjee:2013wna,Nath:2016kfp}, where limits are set for benchmark spectra.
In contrast, we will develop a strategy custom-built for the compressed gluino region. 
Another important difference is in the motivation: while Ref.~\cite{Nath:2016kfp} achieves the correct Higgs mass from radiative corrections and carefully analyzes dark matter constraints, we ignore these considerations, focusing only on our simplified spectrum and its collider strategy.
Finally, specialized analyses similar in spirit to the strategy proposed here but aimed at compressed electroweakino spectra have been studied \cite{Giudice:2010wb, Han:2013usa,Schwaller:2013baa, Baer:2014cua,Han:2014kaa,Bramante:2014dza,Han:2014xoa,Baer:2014kya, Bramante:2014tba,Han:2015lma, Bramante:2015una,Ismail:2016zby}. 

\section{The Analysis}
\label{sec:analysis}

At the LHC, following QCD pair-production, gluinos decay to the LSP and jets through off-shell squarks: 
\beq
p~p \ra \sg~\sg \ra 2 (\sq)^*+2 j \ra 2\dm + 4 j~.
\eeq

We consider in our analysis only decays to the first two quark generations, taken massless.
To begin with, it is enlightening to study the partonic level decay,
$\sg_1 \sg_2 \ra (\dm q \bar{q})_1(\dm q \bar{q})_2$.
The final state jets then have the following two features:
\\

\textit{Partonic Feature A}

 The invariant mass of a pair of quarks that each gluino decays into is bounded from above by the mass splitting between the gluino and the LSP, i.e.,
\beq
m_{(q\bar{q})_i} \leq \Delta M,~~~i=1,2~.
\label{eq:partoninvmasslimit}
\eeq
\\

\textit{Partonic Feature B}

Heavy gluinos are produced nearly at rest in the lab frame, leaving the decay products with little energy to carry.
The jets produced in such events then tend to be soft:

\beq
E_{q_i} \lsim O (\Delta M),~~~i=1,2,3,4~.
\label{eq:partonenergyedge}
\eeq
The exact upper bound on the energy depends on $\hat{s}$ and the mass of the gluino being produced, and can be determined by locating the edge of the energy distributions of the individual jets.
For example, at $\shat =$ 13 TeV, we find the maximum jet energy to be 220 GeV for a 600 GeV gluino (and a 500 GeV LSP).
This edge occurs at smaller energies for heavier gluinos and for $\shat =$ 8 TeV.
This feature can also be interpreted in terms of the angular separation between the $q$-$\bar{q}$ pairs, $\Delta\theta_{q\bar{q}}$. 
For small angles, we have $m_{q\bar{q}} \simeq (E_q E_{\bar{q}})^{1/2}\Delta\theta_{q\bar{q}}$.
If a cut $E_{\rm min}$ were imposed as on the energy of the jets, the angular separation would be bounded on both sides: 
\beq
\frac{m_{q\bar{q}}}{\delm} \lsim \Delta\theta_{q\bar{q}}  \leq \frac{m_{q\bar{q}}}{E_{\rm min}}~.
\label{eq:partonangsep}
\eeq
On the other hand, the angular separation of background jet pairs is not bounded from below since Eq.~(\ref{eq:partonenergyedge}) does not apply to them.

As we now move to more realistic, hadronic level events, the picture changes in two important respects.
Firstly, the event is generally contaminated with ISR. 
As mentioned in the Introduction, monojet + MET searches take advantage of this feature, even though the sensitivity may only improve marginally, as shown by the 13 TeV, high-luminosity projection made in \cite{Low:2014cba}.
Secondly, due to ISR and FSR effects, the jet multiplcity is generally larger than 4.
Accounting for these complications, the key points of information in the signal event are listed below.
These shall be the features we hope to recover in a realistic analysis.

\begin{itemize}[label={}]

\item(i) As a consequence of Eq.~(\ref{eq:partoninvmasslimit}), at least two different pairs of jets have an invariant mass bounded by $\delm$.

(ii) As a consequence of Eq.~(\ref{eq:partonenergyedge}), the non-ISR jets are generally soft (and much softer than the ISR jets).

(iii) The gluinos, being heavier than 600 GeV, will be slow-moving after being produced\footnote{The assumption of a slow-moving LSP in the lab frame can be easily checked as follows. 
The $\met$ distribution must peak at $\delm$, hence the boost factor $\beta_T\gamma_T = \met/\mdm \approx \delm/\mdm$. Since we consider $\mdm > 500$~GeV, $\beta_T\gamma_T < 0.2$ or $\beta_T < 0.2$.}.
A heavy $\dm$ emitted from a gluino is nearly at rest in the gluino rest frame; therefore, we expect a common boost factor for $\sg$ and $\dm$ when boosting their momenta to the lab frame.
Thus we expect $p_{T_{\sg}}/\met = \msg/\mdm$. 
However, as the gluinos recoil against ISR jets in the event, we have $p_{T_{\sg}} \simeq p_{T_{\rm ISR}} $.
Combining these observations, we see that the distribution of the ratio $p_{T_{\rm ISR}}/\met$ peaks at $\msg/\mdm \simeq 1$ for the signal.
This feature was employed in increasing the sensitivity of compressed top squark searches  by \cite{Hagiwara:2013tva,An:2015uwa,Cheng:2016mcw}, and can be translated to our scenario. 
\end{itemize}

At this point it would be most useful to tag  ISR jets in order to proceed with our analysis.
While this cannot be done exactly, we propose a rudimentary ISR tag based on the energy of the jet;
motivated by Eq.~(\ref{eq:partonenergyedge}), we pick $\Etag = \delm$ as a demarcating value, such that 
a hard jet with energy $> \Etag$ is tagged as an ISR jet, while a jet with energy softer than $\Etag$ is tagged as an ``honest jet".
This choice of $\Etag$ enjoys an advantage in defining our cuts, as will be explained later.
Having distinguished between jets in this manner, we can infer a fourth feature of the signal event:
\begin{itemize}[label={}]
\item(iv) ISR jets are generally expected to be fewer in number than honest jets. 
Put differently, the distribution of the ratio $N_{\rm ISR}/N_{\rm honest}$ is expected to peak at lower values for the signal compared to the background.
\end{itemize}

\subsection{Backgrounds}
\label{subsec:bgs}

The dominant irreducible background to our process is $Z+4j$ production followed by the invisible decay of $Z$ to neutrinos. 
Another large background is also generated by processes with ``lost leptons": $W+4j$, $t\bar{t}$ and single-top production can result in jets, MET and a lepton, and the lepton may be lost due to failed isolation, reconstruction or acceptance.
This lost lepton background turns out to be more important than $Z+4j$.
When our basic cuts (to be described in the next sub-section) and vetos of charged leptons and $b$-jets are applied, the background cross-sections are as follows\footnote{These values were generated using Madgraph5-aMC@NLO \cite{Alwall:2014hca} at lowest order and using the default parton distribution functions and scale choices.} -- $Z+4j$: 1870 fb, $W+4j$: 1125 fb, $t\bar{t}$: 1764 fb, single top: 316 fb.
The large rates in the lost lepton background are due to the efficiencies of detector simulation; we find about 30\% of leptons are lost and 30\% of $b$-jets fail to be tagged.

These background processes provide multiple jets that are generally more energetic than the honest jets of the signal. 
The jets are also roughly isotropic, unlike the signal jets where most of the soft jets recoil against the hard ISR.
These features can help discriminate the signal from background.

One other potentially dangerous background is QCD multijet production, where $\slashed E_T$ arises from mismeasured jets.
The prodigious rates of these processes (about $10^9$ pb) precludes their estimation by Monte Carlo simulation.
However, it is possible to render the multijet background harmless with a sufficiently hard $\slashed E_T$ cut, as shown in existing analyses with similar final states, such as 
\cite{Feldman:2009zc,Bramante:2011xd}.
We will assume this approach, and will present in Sec.~\ref{subsec:results} similar cuts to eliminate the QCD background.
See also \cite{Aad:2014wea}, where this background is controlled with cuts on variables such as $\met/\sqrt{H_T}$, where $H_T$ is the sum of jet $p_T$'s passing a selection cut.

\begin{table*}
\begin{center}
\begin{tabular}{|c|c|c|c|}
\hline
Cut & Signal cross-section (fb) & $Z + 4j$ cross-section (fb) & ``Lost leptons" cross-section (fb) 
\\
\hline \hline
Basic cut + trigger & $5.77 \pm 0.06 $ & $1390 \pm 13 $  & $2282 \pm 46$  \\ \hline
Cut I & $3.05 \pm 0.04$  & $393 \pm 7$ & $544 \pm 22$  \\ 
 & (53\%) & (28\%)& (24\%)  \\ \hline
Cut II & $2.72 \pm 0.04$  & $288 \pm 6 $& $393 \pm 18$ \\ 
 & (47\%) & (21\%)& (17\%)  \\ \hline
Cut III & $2.24 \pm 0.04$  & $145 \pm 4 $& $242 \pm 15$  \\ 
 & (39\%) & (10\%) & (10\%) \\ 
\hline
\end{tabular}
\end{center}
\caption{Signal and background leading order cross-sections at the end of each cut, for $\msg = 1$~TeV and $\meteight = 60$~GeV at $\sqrt{s} = 8$~TeV, $\mathcal{L} = 20~\rm{fb}^{-1}$.
The errors shown are statistical.
The efficiency with respect to the original cross-section is denoted in parentheses.
Cut III can be seen to be the strongest discriminator.
The cuts are described in the text.}
\label{tab:cutflow}
\end{table*}

\subsection{Cut Flow}
\label{subsec:cutflow}

We generate partonic level events with Madgraph5-aMC@NLO \cite{Alwall:2014hca}, demanding jet $p_T>$~40 GeV and $|\eta| < 2.5$, and $\met > 60~$GeV (90 GeV) for $\sqrt{s}$ = 8 TeV (13 TeV).
The set of parton distribution functions (PDFs) used is NN23LO1 \cite{Ball:2012cx}.
We then feed the events to PYTHIA6.420 \cite{Sjostrand:2006za} for parton showering, and simulate the detector response with PGS4 \cite{PGS}.
The jets are identified by an anti-$k_T$ jet-clustering algorithm \cite{Cacciari:2008gp} with distance parameter 0.5.
The $b$-tagging efficiency peaks at 50\% with a mistag rate of $1$-$2\%$.
For a description of the reconstruction of $b$-jets, taus, leptons and photons, see \cite{PGS}.

For our analysis, we first veto events with $b$-tagged jets, hadronically decaying taus, photons and leptons, then apply basic cuts requiring a minimum jet $p_T$ of 40 GeV, and $\met > \meteight (\metthirt)$ for $\sqrt{s} = 8$ TeV (13 TeV).
The values of $\meteight$ and $\metthirt$ chosen are partly set by triggering requirements and partly by ambiguities in the background estimation. 
 
To make a realistic estimate of the signal significance, we take into account the full working range of existing triggers.
For this we use the simulated trigger efficiencies provided in \cite{TrigATLAS8TeV} for $\shat$ = 8 TeV and \cite{TrigCMS13TeV} for $\shat$ = 13 TeV.
The 8 TeV trigger (called ``EF $E_T^{\rm miss}>60$~GeV, $L2>40$~GeV, $L1>$~35 GeV") turns on at $\met = 50$~GeV with 10\% efficiency and steadily improves until it reaches 100\% efficiency at $\met = 140$~GeV. 
It remains fully efficient until $\met = 300~$GeV, beyond which the efficiencies are not presented.
The MET bin size in this plot is 10 GeV.
Similarly the 13 TeV trigger (``\url{HLT_PFMETNoMu90_JetIdCleaned_PFMHTNoMu90_IDTight}")  turns on at $\met =$ 90 GeV at 17\% and becomes 100\% efficient in the range 270-460 GeV, with bin size 10 GeV.
To apply these triggers,
we first make $\met$ distributions of the signal and background with the corresponding bin size and range, and then convolve them with the efficiencies.
It is interesting to consider the gain or loss by accounting for the full trigger range and its efficiencies as just described.
Suppose first a hypothetical scenario in which the trigger is ideal, i.e., it is 100\% efficient in its entire working range. 
Then our procedure of convolving MET distributions with trigger efficiencies reduces to simply using the former to discriminate signal and background. 
In that case, we find the significance to improve by at most 10\%.
Therefore, the use of imperfect efficiencies results in only a small loss in significance for compressed spectra. 
Next, let us consider the outcome if we had ignored imperfect efficiencies, and only selected events from the fully efficient region. 
Now the significance declines by a factor of $\sim$ 1.5.
LHC searches utilize, unfortunately, only this limited trigger range where the efficiency is 100\%.
The signal acceptance of compressed spectra is further impaired in these searches by the imposition of a hard cut on the scalar sum of jet $p_T$'s ($H_T$).
Due to these practices, the signal significance is greatly reduced.
\\

We now construct a three-step cut flow taking into account the signal features (i)-(iv) mentioned in the beginning of the section.
\\

\textbf{Cut I} 

In the absence of ISR, the signal event must contain multiple soft jets.
We demand 
\beq
N_{\rm honest} \geq n~.
\eeq
The invariant mass squared of a jet pair is $m_{j_1 j_2}^2= 4 E_1 E_2 \sin^2(\Delta\theta_{12}/2) $. 
We expect the angular separation between the honest jets to be small, since they are emitted from near-collinear gluinos recoiling against ISR.
Thus,  $m_{j_1 j_2}^2\approx E_1 E_2 \Delta\theta^2_{12} $.
By choosing $\Etag = \delm$, we  ensure $m_{j_1 j_2} \leq \delm$.
It follows that if we pick $n$ = 4, this cut is tantamount to the statement in signal feature (i).
We therefore use this as our choice of $n$. Smaller values of $n$ would admit more signal, but the background is also much larger.
\\

\textbf{Cut II}

As the gluinos and ISR jets are back-to-back, we ask the MET and the $p_T$ of the hardest ISR jet to be on opposite sides of the beam axis: 
\beq
||\Delta \phi (\met, j_{\rm ISR, max})| - \pi | \leq 1.5~.
\eeq
\\
\\
\\

\textbf{Cut III}

Taking into account features (ii)-(iv), we impose
\beq
\newvar \equiv \frac{\sum^{N_{\rm ISR}}_{i=0}E^i_{\rm ISR}}{\met} \frac{N_{\rm ISR}}{N_{\rm honest}} \leq k(\sqrt{s},\msg),
\label{eq:Rdefn}
\eeq
where $E^i_{\rm ISR}$ are the energies of ISR jets. 
$k(\sqrt{s},\msg)$ is an $\mathcal{O}(1)$ number optimized for the collider center-of-mass energy $\sqrt{s}$ and for the gluino mass.
In practice, we find $k = 0.9-2$ for $\sqrt{s} = 8$~TeV and 
$k = 2$ for $\sqrt{s} = 13$~TeV.
The primary difference between the cut used in \cite{Hagiwara:2013tva,An:2015uwa,Cheng:2016mcw} for compressed stop searches and our cut here is the weighting by the ratio $N_{\rm ISR}/N_{\rm honest}$ in our case.
This a reflection of signal feature (iv), which is unique to gluino production -- unlike squark production, the final state here (not counting the ISR) must contain four jets. 
Due to this weighting, we expect the signal events to occupy the $\newvar$ distribution chiefly at values close to zero.
Another difference is our use of ISR jet energies instead of the $p_T$ of the leading jet. 
We find the background distribution of the former more even, leading to a clear peak in $\newvar$ for the signal.

We will find this cut to be the strongest discriminant between signal and background, as it takes into account the uniquest features of the signal: the slowness of $\dm$, the back-to-back nature of the gluinos and ISR, and the multiplicities of ISR and honest jets.
These features are absent in the background, which is more isotropic, and has generally fewer honest jets and more ISR jets.

\begin{figure*}
\begin{center}
\includegraphics[width=.45\textwidth]{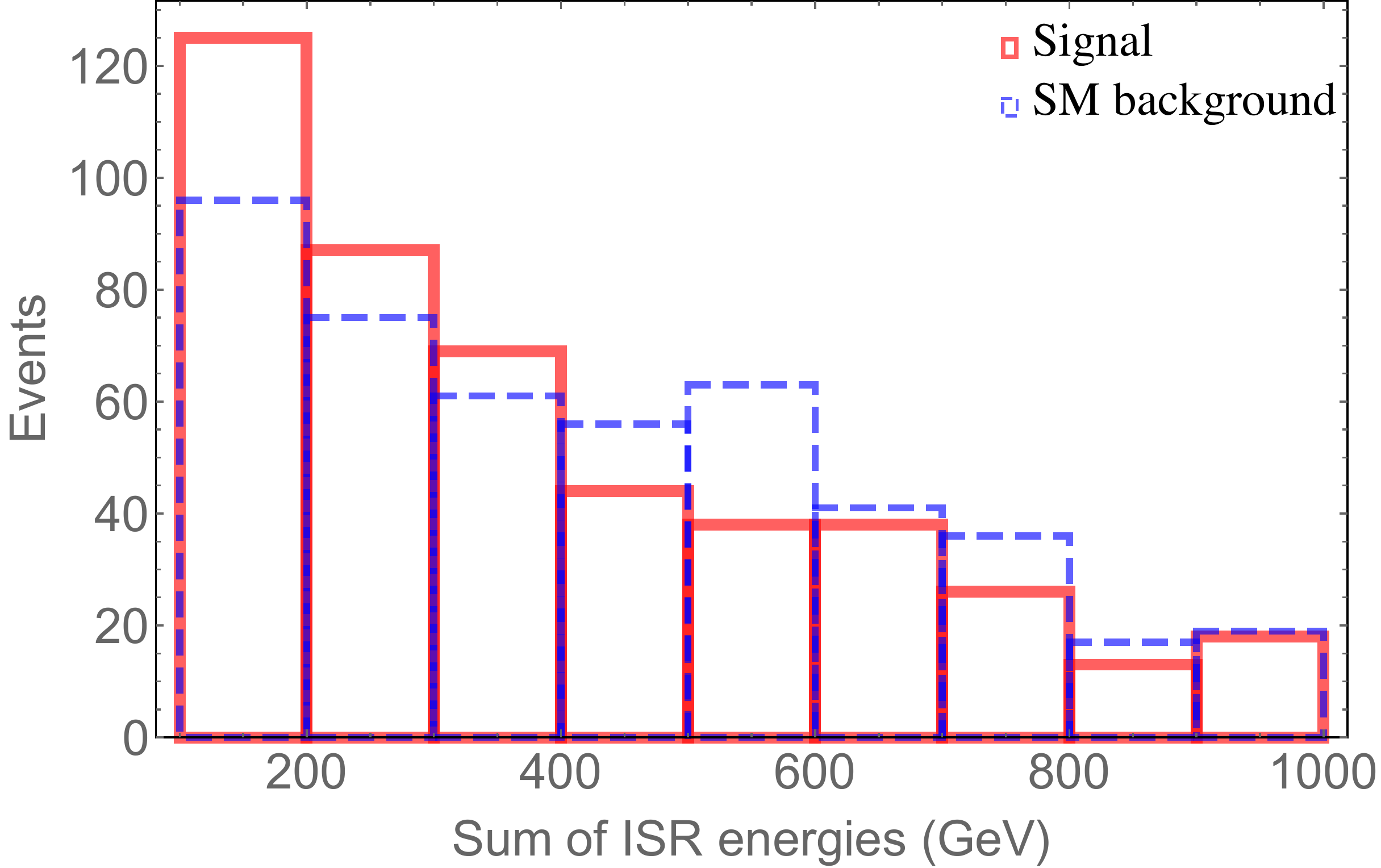} \quad \quad
\includegraphics[width=.45\textwidth]{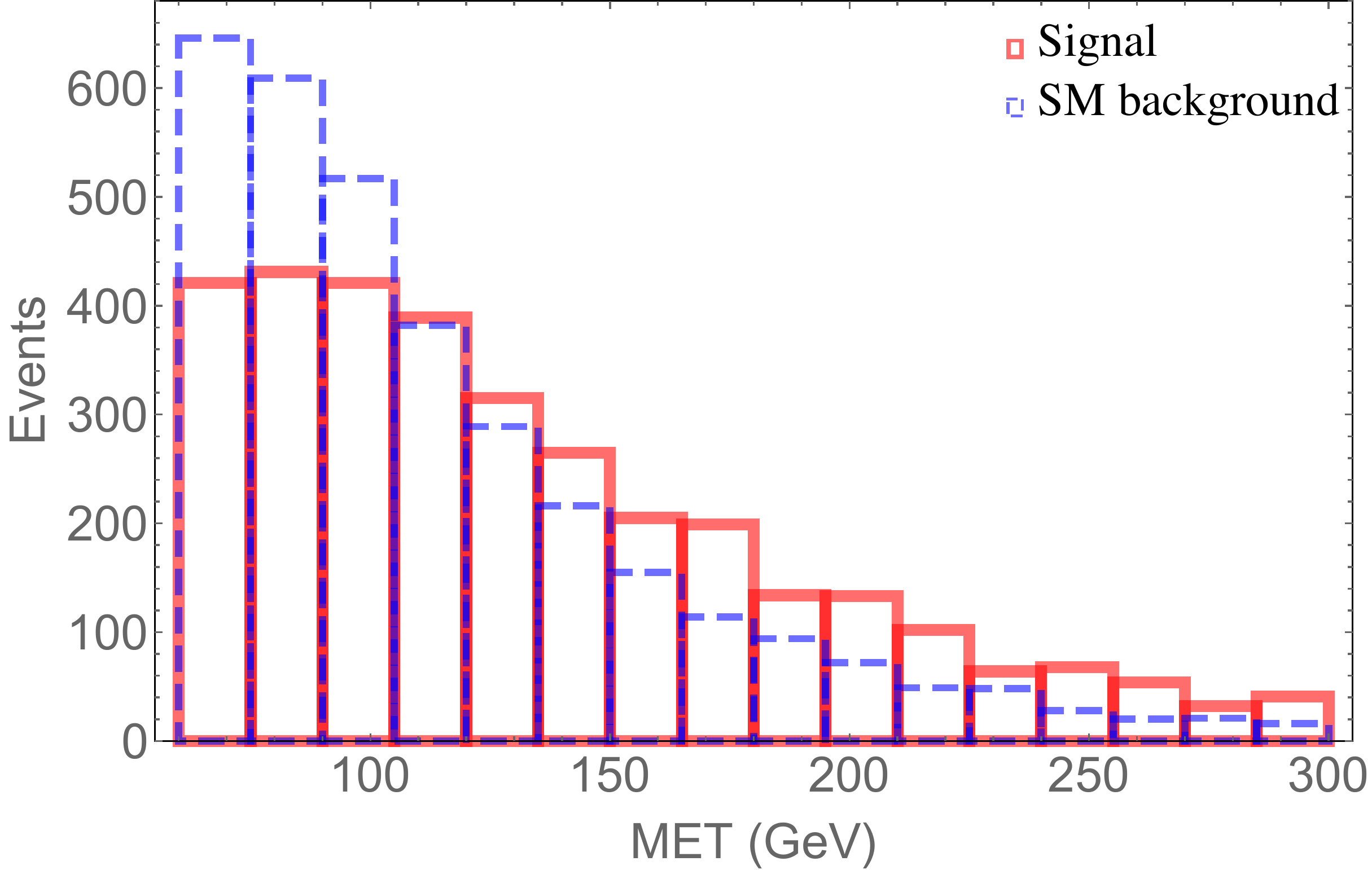} \\ \vspace{.6 cm}
 
\includegraphics[width=.45\textwidth]{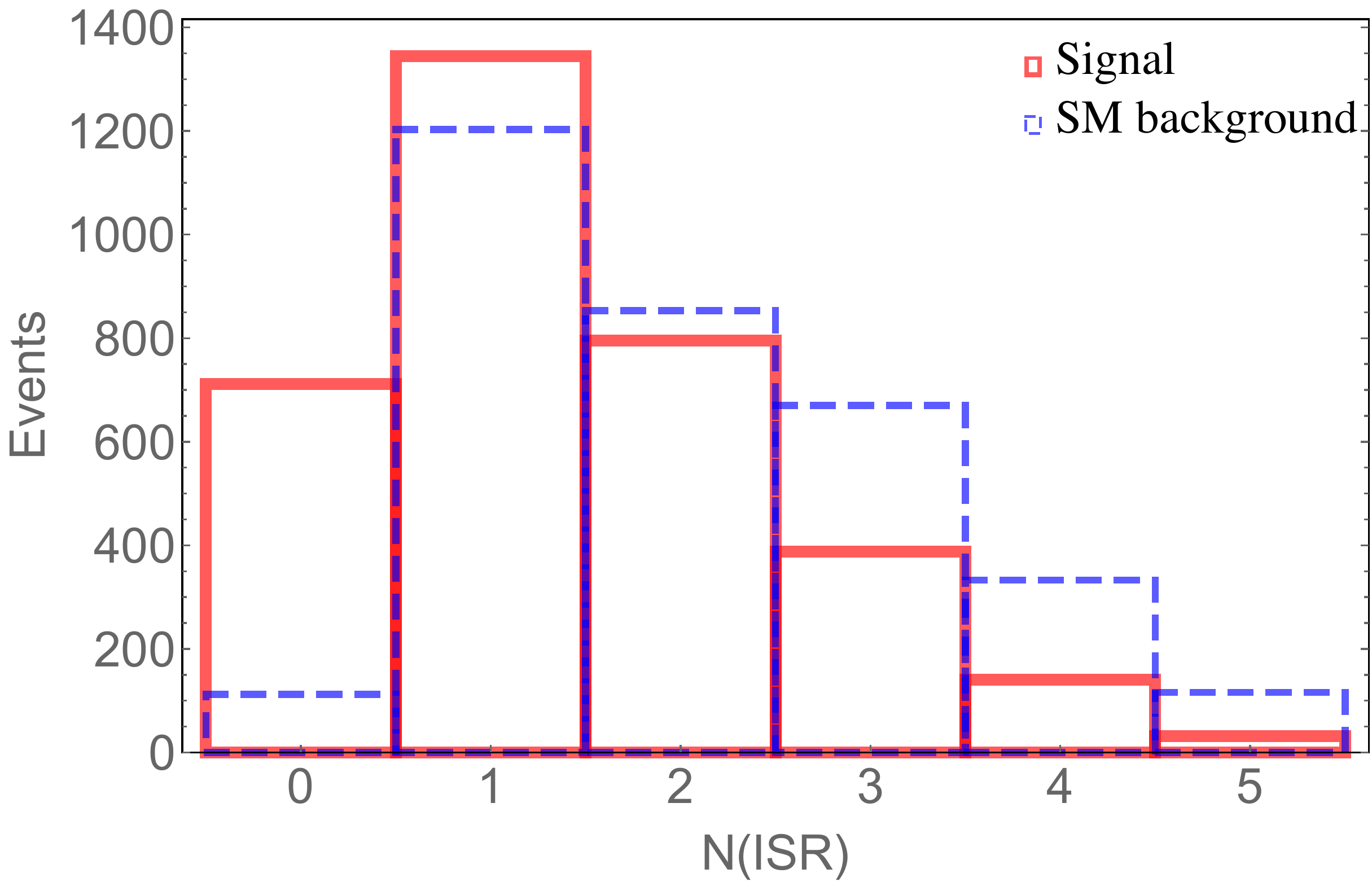} \quad \quad
\includegraphics[width=.45\textwidth]{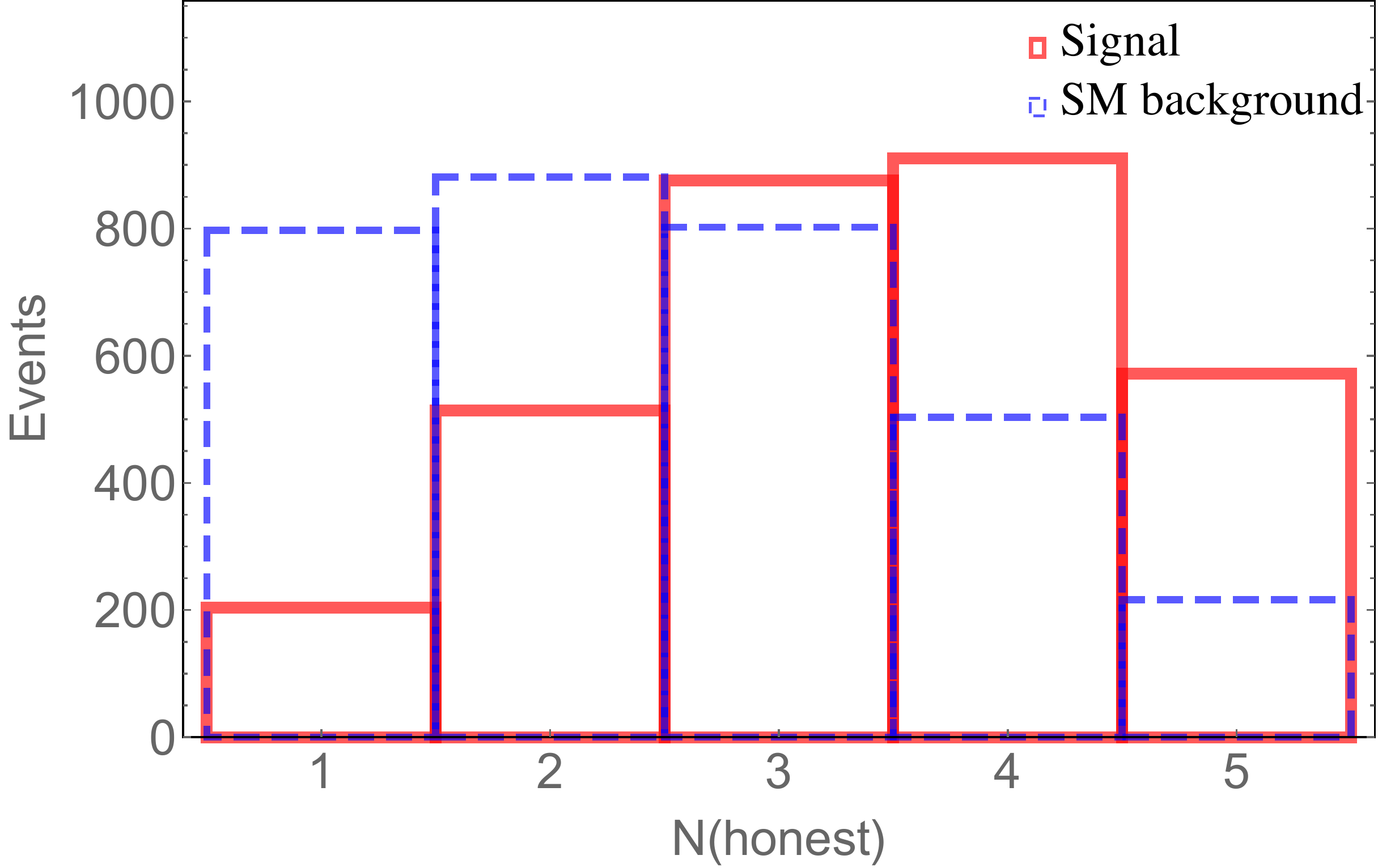}
\caption{Area-normalized event distributions of the quantities on the LHS of Eq.~(\ref{eq:Rdefn}).
We take $\sqrt{s} = 8$~TeV, $\mathcal{L} = 20~\rm{fb}^{-1}$ and $\msg = 1$~TeV for illustration.
From these distributions we can infer that the combination of quantities that defines $\rho$ peaks at smaller values for the signal.
See also Fig.~\ref{fig:moneycut} that validates this.
}
\label{fig:histograms}
\end{center}
\end{figure*}

To further visualize how $\rho$ distinguishes signal from background, we chart in Fig.~\ref{fig:histograms} area-normalized histograms of each quantity on the LHS of Eq.~\ref{eq:Rdefn}.
For illustration we take for the signal $\msg = 1$~TeV, and generate events at $\sqrt{s} = 8$~TeV.
We find that for the relative positions of the signal peak with respect to background are the following for each quantity -- $\sum^{N_{\rm ISR}}_{i=0}E^i_{\rm ISR}$: lower, $\met$: higher, $N_{\rm ISR}$: lower, $N_{\rm higher}$: higher. 
The overall effect is to tug more signal than background toward smaller $\rho$.

\begin{figure}
\begin{center}
\includegraphics[width=8cm]{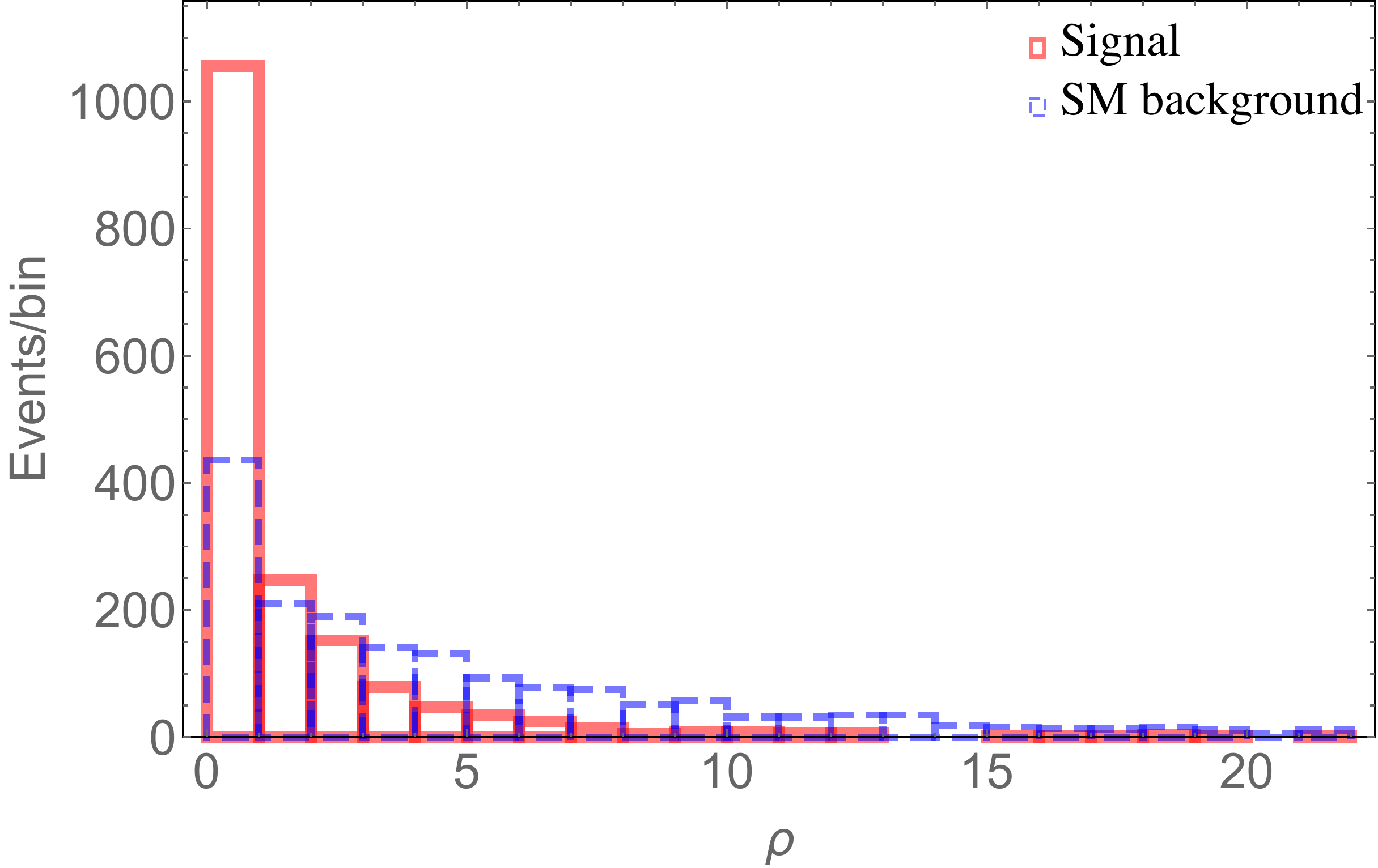}
\caption{Area-normalized event distributions in $\newvar$ (defined in Eq.~(\ref{eq:Rdefn})) after imposing Cuts I and II.
We take $\sqrt{s} = 8$~TeV, $\mathcal{L} = 20~\rm{fb}^{-1}$ and $\msg = 1$~TeV for illustration.
The signal distinctly peaks near low $\newvar$ because of kinematic features (ii)-(iv) described in the text.
}
\label{fig:moneycut}
\end{center}
\end{figure}

\subsection{Results}
\label{subsec:results}

Before presenting the results of our analysis, we will clarify some subtleties regarding our background estimation.

 First, mismeasured jets from QCD production provide a background component that cannot be simulated due to their high rates.
This problem is usually confronted by applying certain cuts on jet multiplicity, jet $p_T$, $\met$ and the angular separation between the leading jets and $\met$, and proceeding under the assumption that this background is removed \cite{Bramante:2011xd}.
To our knowledge, there is no consensus on what the choices of cuts are (they must be derived from data and depend on the final state). 
Therefore, multiple values are considered for our basic $\met$ cuts:
\begin{equation*}
\meteight =  60,~100,~140~{\rm GeV};~~~\metthirt =  90,~180~{\rm GeV}~. 
\end{equation*}
The lowest of these values is chosen following existing triggers
and gives the most optimistic bound or discovery reach.
We assume that by virtue of these cuts, alongside our basic jet $p_T$ cut and large jet multiplicity, the QCD multijet background is eliminated.

Next, we have used lowest order values for all cross sections and approximated the $W/Z + \text{jets}$ backgrounds with $W/Z+4\,~ \text{jet}$. Higher order processes and the uncertainty on the background cross section will be subsumed into a multiplicative factor. One may worry that a flat multiplicative factor does not sufficiently capture the effects of higher jet multiplicity processes, as $W/Z + 5^+\, \text{jets}$ have more jets and should more easily pass the basic selection cuts.
However, higher multiplicity processes also have more jets passing our ``ISR jet" criteria and thus tend to have higher values of $\rho$, resulting in a lower final efficiency. To test the impact of higher multiplicity processes on our background estimates, we compared a $Z+4\, \text{jet}$ sample with a matched $Z + 4^+\, \text{jet}$, both generated with SHERPA 2.2.0~\cite{Gleisberg:2008ta}. Passing both background estimates through our analysis, we find the final efficiencies are comparable, well within the scale and parton distribution function uncertainties.
Therefore we will adhere to the simpler procedure of modeling the full background with lowest order $W/Z+4\,~ \text{jets}$ and a multiplicative $O(1)$ uncertainty factor.\\

Table~\ref{tab:cutflow} shows the efficiencies of the cut flow employed in this paper.
For illustration, we take $\msg = 1$~TeV, $\meteight = 60$~GeV at $\sqrt{s} = 8$~TeV, $\mathcal{L} = 20~\rm{fb}^{-1}$.
Cuts I and II are already capable of selecting more signal events than the background, but we see that it is Cut III that makes the greatest difference.
As further illustration of this point, Fig.~\ref{fig:moneycut} shows the signal and background distributions in the variable $\newvar$ after the imposition of Cuts I and II.
We see a clear peak in the signal distribution at low $\newvar$, as opposed to the background which is more evenly distributed.
As explained earlier, the ratio $N_{\rm ISR}/N_{\rm honest}$ helps populate the signal more densely near $\newvar = 0$.

After all the cuts are imposed, we obtain the significance of the signal defined as 
\beq
\sigma \equiv \frac{S}{\sqrt{B(1+\delta)}}~,
\label{eq:sigmadef}
\eeq
with $S$ and $B$ the number of signal and background events and where $\delta$ captures systematic uncertainties in the background estimate that we will describe shortly.

\begin{table*}
\begin{center}
\begin{tabular}{c | c   c}
\multicolumn{3}{c}{$\sqrt{s}$ = 8 TeV, $\mathcal{L} = 20~\rm{fb}^{-1}$} \\
\hline
$\meteight$ & $3\sigma$ & $5\sigma$ 
\\
\hline \hline
60 GeV & 900 GeV & 850 GeV  \\ \hline
100 GeV & 890 GeV & 840 GeV  \\ \hline
140 GeV & 880 GeV & 825 GeV \\ \hline
\end{tabular}
\quad \quad \quad \quad
\begin{tabular}{c|c c}
\multicolumn{3}{c}{$\sqrt{s}$ = 13 TeV, $5\sigma$ reach} \\
\hline
$\metthirt$ & $\mathcal{L} = 20~\rm{fb}^{-1}$  & $\mathcal{L} = 3~\rm{ab}^{-1}$ 
\\
\hline \hline
90 GeV  & 990 GeV & 1370 GeV  \\ \hline
180 GeV & 980 GeV &  1360 GeV  \\ \hline
\end{tabular}
\end{center}

\caption{Summary of our results. 
These limits are quoted using the central values in curves such as Fig.~\ref{fig:Sigmas}, given the uncertainties in background estimation. 
 The QCD multijet background is assumed eliminated for each value of $\meteight$ and $\metthirt$ shown.
}
\label{tab:results}
\end{table*}

\begin{figure*}
\begin{center}
\includegraphics[width=.45\textwidth]{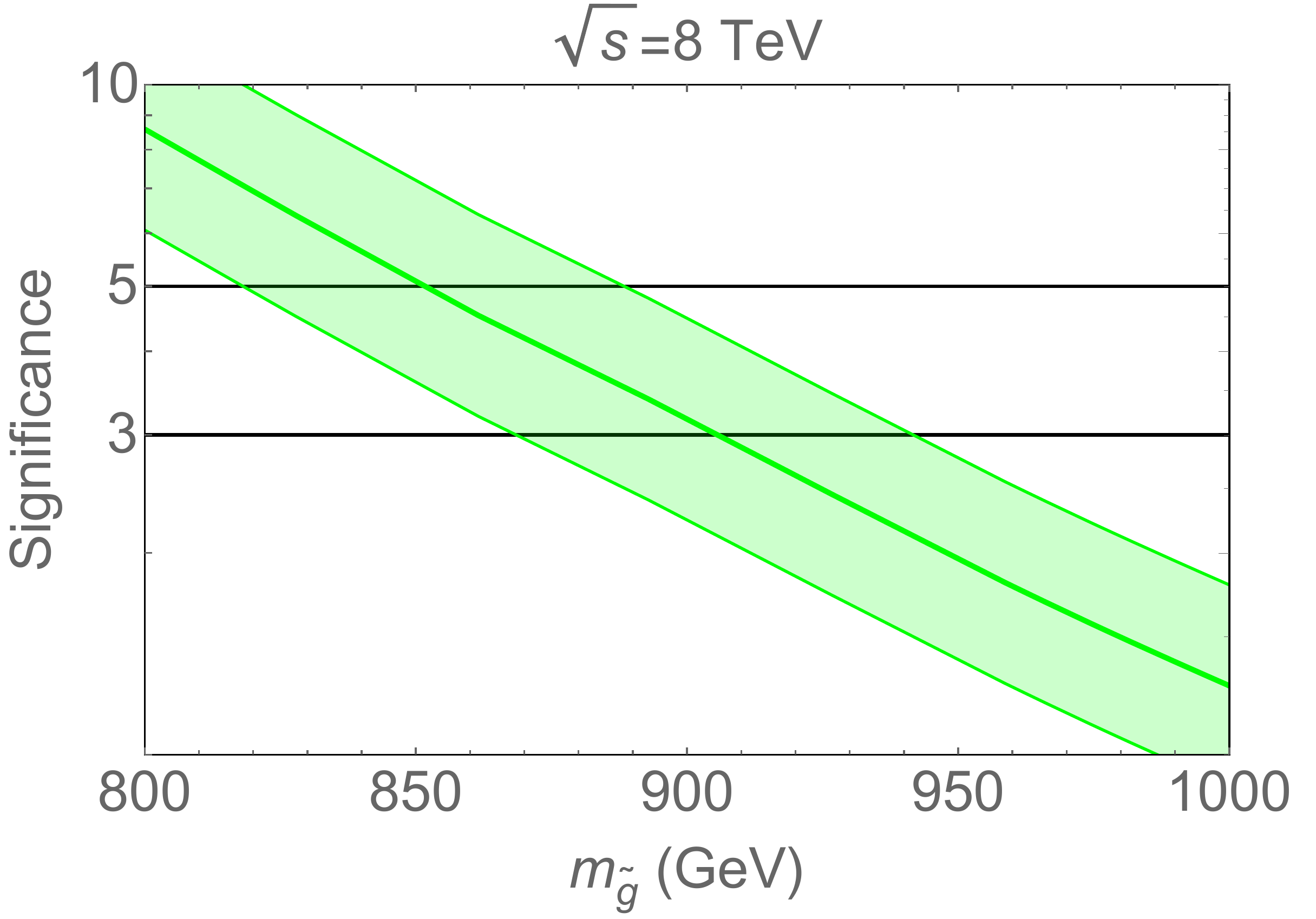}
\includegraphics[width=.45\textwidth]{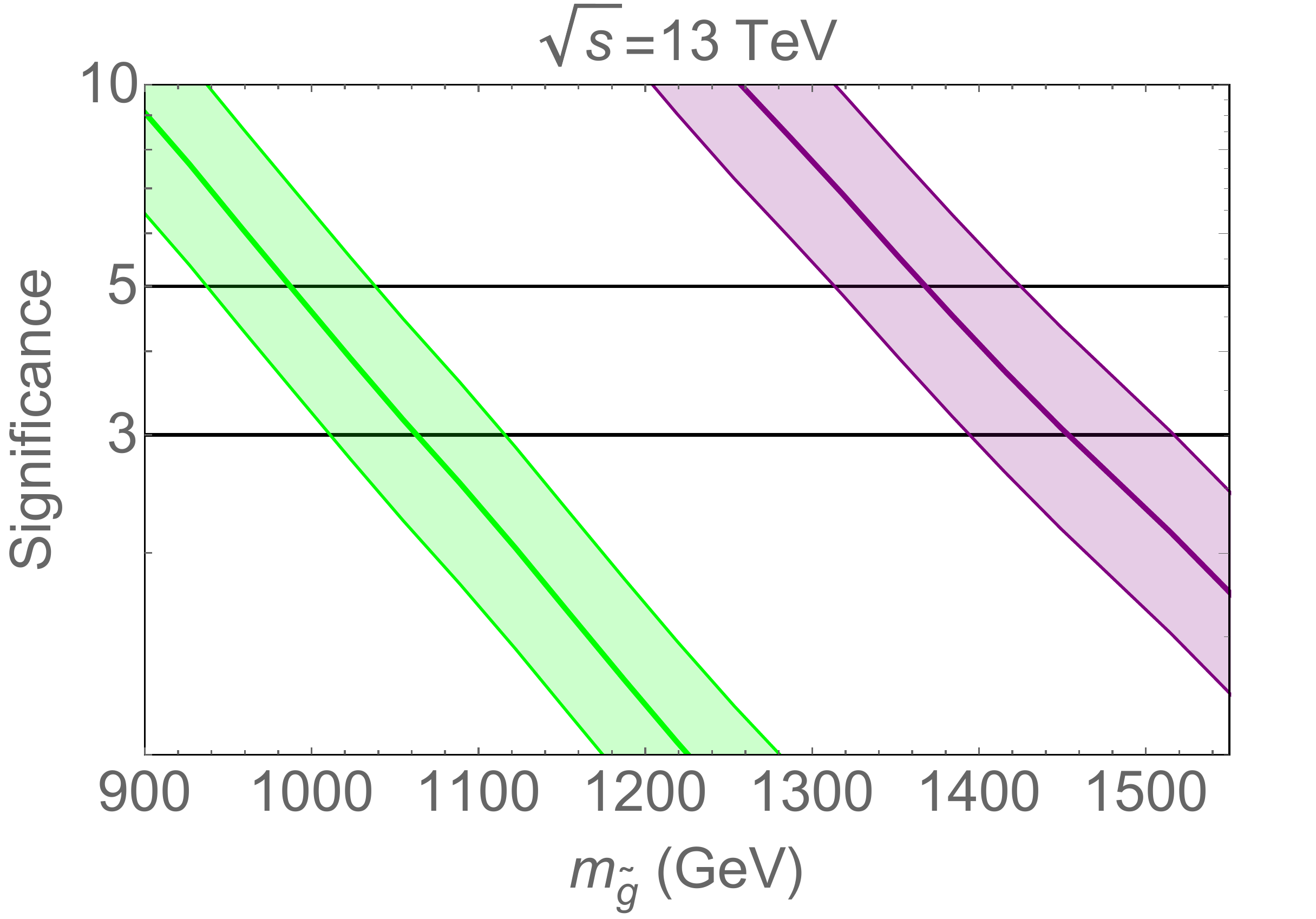}
\caption{Significances (as defined in Eq.~(\ref{eq:sigmadef})) of the compressed gluino search after all our cuts are imposed.
The plot on the left (right) corresponds to $\sqrt{s} = 8 (13)$~TeV.
The green (purple) curves denote an integrated luminosity of $20~\rm{fb}^{-1} (3~\rm{ab}^{-1})$.
The bands depict background systematic uncertainties in the range [-50\%,100\%] due to matching procedures and factorization scale choice. 
}
\label{fig:Sigmas}
\end{center}
\end{figure*}

In Fig.~\ref{fig:Sigmas} we plot $\sigma$ as a function of $\msg$ after passing the events through our cuts.
The left-hand plot corresponds to $\sqrt{s}=8$~TeV at an integrated luminosity $\mathcal{L} = 20~\rm{fb}^{-1}$, with $\meteight = 60$~GeV.
The right-hand plot shows our projected sensitivity to searches at $\sqrt{s} = 13$~TeV, the green (blue) curve corresponding to $\mathcal{L} = 20~\rm{fb}^{-1} (3~\rm{ab}^{-1})$.
Here, $\metthirt = 90$~GeV.
In making these plots, we vary the background by a factor of 2 in either direction to allow for uncertainties due to matching procedures and choice of factorization scale. 
This amounts to varying the systematic uncertainty $\delta$  between -50\% and 100\%. 
We denote this uncertainty by bands around the curves.
These uncertainties are a conservative choice. For instance, one may compare this with the background uncertainty from the variation of the common renormalization and factorization scale up and down by a factor of 2. 
The background cross-section at 8 TeV then varies by a maximum of $27\%$, and at 13 TeV by a maximum of $26 \%$.
Similarly, one may wish to compare with the PDF uncertainty. 
If we use the PDF set CTEQ6L1, the background cross-section at 8 TeV (13 TeV) changes by only 12\% (7\%).

We reassert that these plots represent cuts that assume that the QCD multi-jet background is removed. 
Analogous curves may be drawn for other choices of $\meteight$ and $\metthirt$; which choice is best-suited for eliminating the background can only be verified by experiment.

If we take the central values of the bands in Fig.~\ref{fig:Sigmas}, we get the following results. 
First, even with current 8 TeV data, we have a $5\sigma$ discovery reach of out to about $\msg = 850$~GeV. 
The $3\sigma$ exclusion limit is $\msg = 900$~GeV.
This is our most optimistic result, based on the assumption that the QCD multijet background is removed for $\meteight = 60$~GeV.
We expect the limits to be looser for harder cuts on the MET.
Therefore, for $\meteight = 100$~GeV (140 GeV)
the 5$\sigma$ reach falls to $\msg = 840$~GeV (825 GeV) and the 3$\sigma$ limit to $\msg = 890$~GeV (880 GeV).
This is the principal result of our paper.
We remind the reader that current search techniques based on jets + MET and MT2-like variables have no sensitivity at $\msg > 600$~GeV for gluinos in the compressed region.
As we go to the 13 TeV LHC, in the optimistic case of $\metthirt = 90$~GeV, the $5 \sigma$ reach extends to $\msg$ = 990~GeV with a luminosity of $20~\rm{fb}^{-1} $.
If we ramp up the luminosity to $3~\rm{ab}^{-1}$, we can push the discovery reach to 1370~GeV.
 As before, a harder MET cut affects this result slightly.
 For $\metthirt = 180$~GeV, the reach is $\msg$ = 980~GeV (1360~GeV) for a luminosity of $20~\rm{fb}^{-1} (3~\rm{ab}^{-1})$.
These results are summarized in Table~\ref{tab:results}.
 
We summarize with the two reasons why we get stronger limits and higher sensitivies.  
First, we devised cuts that exploit the unique kinematic features of the compressed gluino-bino spectrum.
In particular, the $\newvar$ variable takes advantage of the recoil of neutralinos and jets against ISR, and of the large multiplicity of soft jets in the signal. 
Second, we made full use of available triggers.
Current searches at ATLAS and CMS select signal regions where the trigger is $100 \%$ efficient. 
This usually results in hard cuts on jet $p_T$ (or total $H_T$) and $\met$, killing sensitivity to compressed regions.
We have shown how softer cuts can be imposed by engaging the entire range of the trigger despite imperfect efficiencies. 
While our analysis may not be completely realistic, we have shown that good sensitivities are achievable.
We therefore urge experimental collaborations to make trigger menus more publicly available.
\\

 \section{Discussions}
 \label{sec:fin}

In our work we took a fixed mass splitting, $\delm = 100~$GeV. 
The variation of this parameter has a non-trivial effect on our strategy.
As the signal MET peaks at $\delm$, for smaller values the background and signal MET distributions begin to overlap. 
Further, the poor trigger efficiencies at small $\met$ filter fewer signal events.
As we dial $\delm$ higher than 100 GeV, the cuts we impose start losing their power as they 
are devised specifically for the features of a compressed region;
for the gluino masses considered, we find that for $\delm \gsim 180~$GeV our strategy is less
sensitive than the usual jets+MET cut-and-count strategy (taking into account the full trigger range).
Therefore, $\delm = 100~$GeV is an optimum value for the method presented here.

\begin{figure}
\begin{center}
\includegraphics[width=8cm]{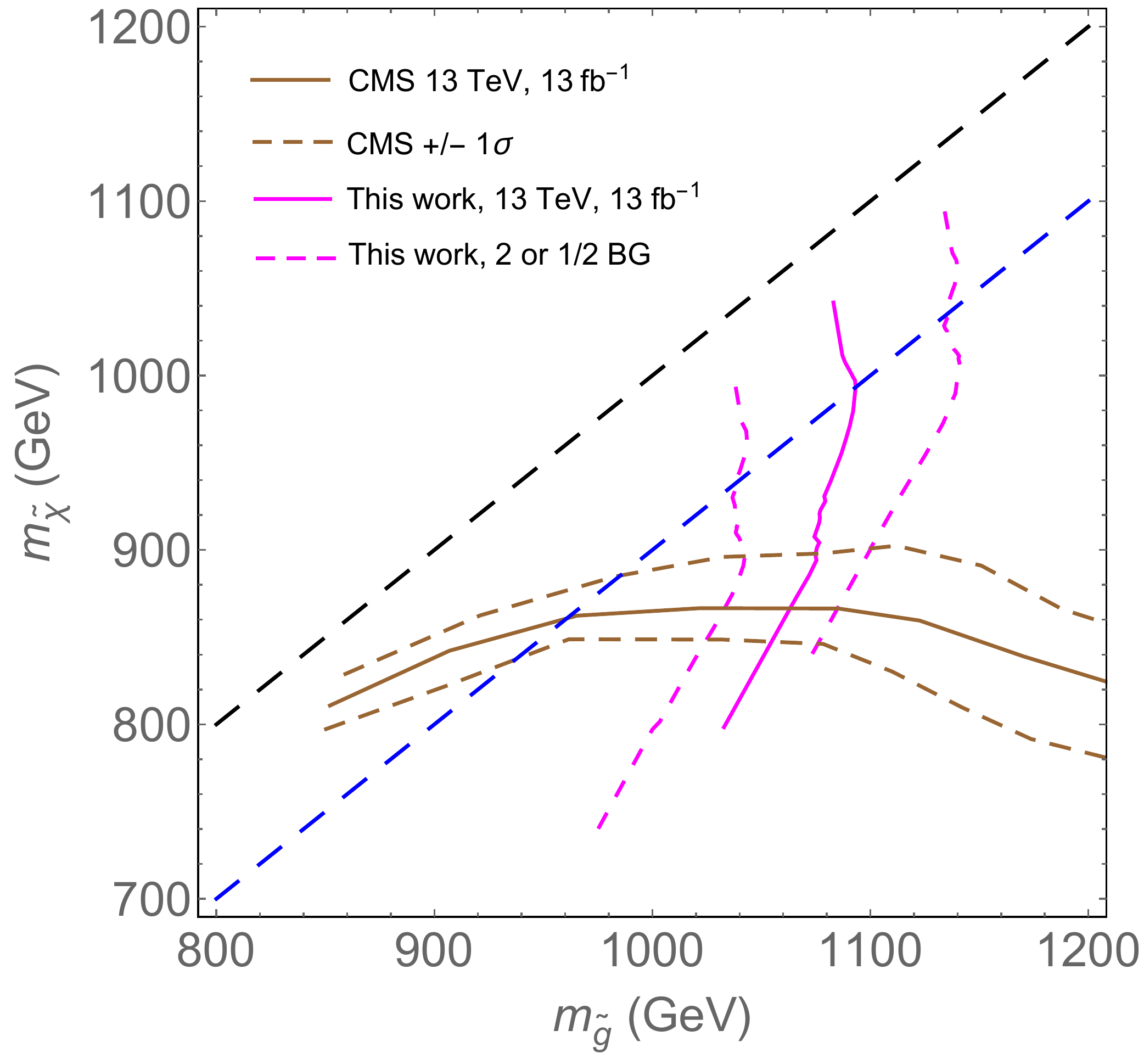}
\caption{95\% exclusion contours in the $\msg$-$\mdm$ plane at $\sqrt{s} = 13$ TeV and $\mathcal{L} = 12.9 \ {\rm fb}^{-1}$. 
The dashed black (blue) line is where $\msg$ = $\mdm$ ($\msg$ = $\mdm$ + 100 GeV.)
The magenta curves depict our strategy, the brown curves are CMS exclusion limits.
This plot illustrates that a dedicated search strategy for the compressed region ($\delm \lsim 200~$GeV) can provide better limits than existing jets+MET-based cut-and-count strategies. 
}
\label{fig:13TeVContour}
\end{center}
\end{figure}
We further illustrate the above point by making an ``apples-to-apples" comparison between our strategy and the most recent search presented by CMS in the Run II of LHC, with $\sqrt{s} = 13$ TeV and $\mathcal{L} = 12.9 \ {\rm fb}^{-1}$ \cite{CMS1313MT2}.
In Fig.~\ref{fig:13TeVContour}, the solid brown curve depicts the 95\% C.L. exclusion limit by CMS in the $\msg$-$\mdm$ plane; the dashed brown curves are the $\pm 1 \sigma$ uncertainties.
The solid magenta curve is the $2\sigma$ exclusion curve of our strategy; the dashed magenta curves capture the uncertainty for background variations by a factor of 2. 
Here we have taken the basic MET cut $\metthirt$ to be $90$ GeV, which is our most optimistic scenario for eliminating the QCD background.
For $\delm = 100$ GeV, CMS excludes $\msg \leq 900$~GeV, whereas our strategy can exclude $\msg \leq 1090$~GeV.
As we increase $\delm$ our strategy worsens -- for $\delm \gsim 200~$GeV, the CMS search clearly overtakes us, as observed in the regions right-side of points where the brown and magenta curves cross.
This plot demonstrates the need for experimental collaborations to use dedicated search strategies in compressed regions.

As already mentioned in the Introduction, we may extend our strategy to compressed spectra involving squarks and non-supersymmetric colored particles.
Extensions to uncolored spectra are also possible.
For instance, instead of an ISR jet, an ISR photon or $W/Z$ may be used instead for the search of a compressed slepton-neutralino or chargino-neutralino spectrum.
While ideas along these lines have been explored in \cite{Schwaller:2013baa, Baer:2014cua,Han:2014kaa,Bramante:2014dza,Han:2014xoa,Baer:2014kya, Bramante:2014tba,Han:2015lma, Bramante:2015una,Ismail:2016zby}, we venture that a variable analogous to $\rho$ may be constructed, involving leptons instead of jets.

If we believe supersymmetry to be the cure to both electroweak fine-tuning and the mystery of dark matter, the null results from conventional searches are only a part of the story. 
The natural place to look next is the compressed region, where light superpartners may hide and co-annihilate with DM to set its abundance.
We urge the theory and experimental communities to take these regions seriously and survey them comprehensively  with dedicated techniques.

\section*{Acknowledgments}

We are grateful to
Joseph Bramante,
Fatemeh Elahi, 
Jared Evans,
Kevin Lannon,
Zhen Liu and
Carlos Wagner
for fruitful conversation, and Jan Winter for SHERPA advice.
This work was partially supported by the National
Science Foundation under Grants 
No. PHY-1417118 and No. PHY-1520966.


\begin{thebibliography}{99}

\bibitem{Griest:1990kh} 
  K.~Griest and D.~Seckel,
  Phys.\ Rev.\ D {\bf 43}, 3191 (1991).
  
\bibitem{Aad:2014nra} 
  G.~Aad {\it et al.} [ATLAS Collaboration],
  Phys.\ Rev.\ D {\bf 90}, no. 5, 052008 (2014)
  [arXiv:1407.0608 [hep-ex]].
  
\bibitem{Khachatryan:2015wza} 
  V.~Khachatryan {\it et al.} [CMS Collaboration],
  JHEP {\bf 1506}, 116 (2015)
  [arXiv:1503.08037 [hep-ex]].
  
  \bibitem{CMSstopcharm} 
  CMS-PAS-SUS-13-009
  
\bibitem{Aad:2015zva} 
  G.~Aad {\it et al.} [ATLAS Collaboration],
  Eur.\ Phys.\ J.\ C {\bf 75}, no. 7, 299 (2015)
  [Eur.\ Phys.\ J.\ C {\bf 75}, no. 9, 408 (2015)]
  [arXiv:1502.01518 [hep-ex]].


  
\bibitem{Low:2014cba} 
  M.~Low and L.~T.~Wang,
  JHEP {\bf 1408}, 161 (2014)
  [arXiv:1404.0682 [hep-ph]].
  
\bibitem{Aad:2014mfk} 
  G.~Aad {\it et al.} [ATLAS Collaboration],
  Phys.\ Rev.\ Lett.\  {\bf 114}, no. 14, 142001 (2015)
  [arXiv:1412.4742 [hep-ex]].
  
  
\bibitem{Hagiwara:2013tva} 
  K.~Hagiwara and T.~Yamada,
  Phys.\ Rev.\ D {\bf 91}, no. 9, 094007 (2015)
  [arXiv:1307.1553 [hep-ph]].
     
\bibitem{An:2015uwa} 
  H.~An and L.~T.~Wang,
  Phys.\ Rev.\ Lett.\  {\bf 115}, 181602 (2015)
  [arXiv:1506.00653 [hep-ph]].
  
\bibitem{Cheng:2016mcw} 
  H.~C.~Cheng, C.~Gao, L.~Li and N.~A.~Neill,
  JHEP {\bf 1605}, 036 (2016)
  [arXiv:1604.00007 [hep-ph]].
  
\bibitem{Macaluso:2015wja} 
  S.~Macaluso, M.~Park, D.~Shih and B.~Tweedie,
  JHEP {\bf 1603}, 151 (2016)
  [arXiv:1506.07885 [hep-ph]].

  
\bibitem{Profumo:2004wk} 
  S.~Profumo and C.~E.~Yaguna,
  Phys.\ Rev.\ D {\bf 69}, 115009 (2004)
  [hep-ph/0402208].
    
\bibitem{Feldman:2009zc} 
  D.~Feldman, Z.~Liu and P.~Nath,
  Phys.\ Rev.\ D {\bf 80}, 015007 (2009)
  [arXiv:0905.1148 [hep-ph]].
  
\bibitem{deSimone:2014pda} 
  A.~De Simone, G.~F.~Giudice and A.~Strumia,
  JHEP {\bf 1406}, 081 (2014)
  [arXiv:1402.6287 [hep-ph]].
  
\bibitem{Harigaya:2014dwa} 
  K.~Harigaya, K.~Kaneta and S.~Matsumoto,
  Phys.\ Rev.\ D {\bf 89}, no. 11, 115021 (2014)
  [arXiv:1403.0715 [hep-ph]].
  
\bibitem{Ellis:2015vaa} 
  J.~Ellis, F.~Luo and K.~A.~Olive,
  JHEP {\bf 1509}, 127 (2015)
  [arXiv:1503.07142 [hep-ph]].
  
\bibitem{Nagata:2015hha} 
  N.~Nagata, H.~Otono and S.~Shirai,
  Phys.\ Lett.\ B {\bf 748}, 24 (2015)
  [arXiv:1504.00504 [hep-ph]].
  
\bibitem{Ellis:2015vna} 
  J.~Ellis, J.~L.~Evans, F.~Luo and K.~A.~Olive,
  JHEP {\bf 1602}, 071 (2016)
  [arXiv:1510.03498 [hep-ph]].
  
\bibitem{Baker:2015qna} 
  M.~J.~Baker {\it et al.},
  JHEP {\bf 1512}, 120 (2015)
  [arXiv:1510.03434 [hep-ph]].
  
  
\bibitem{Aad:2014wea} 
  G.~Aad {\it et al.} [ATLAS Collaboration],
  JHEP {\bf 1409}, 176 (2014)
  doi:10.1007/JHEP09(2014)176
  [arXiv:1405.7875 [hep-ex]].
  
\bibitem{Aad:2015iea} 
  G.~Aad {\it et al.} [ATLAS Collaboration],
  JHEP {\bf 1510}, 054 (2015)
  [arXiv:1507.05525 [hep-ex]].
  
\bibitem{Chatrchyan:2013mys} 
  S.~Chatrchyan {\it et al.} [CMS Collaboration],
  Eur.\ Phys.\ J.\ C {\bf 73}, no. 9, 2568 (2013)
  [arXiv:1303.2985 [hep-ex]].
  
\bibitem{CMS:2014wsa} 
   S.~Chatrchyan {\it et al.} [CMS Collaboration],
  JHEP {\bf 1406}, 055 (2014)
  [arXiv:1402.4770 [hep-ex]].
  
\bibitem{CMS:2014ksa} 
  CMS Collaboration [CMS Collaboration],
  CMS-PAS-SUS-13-019.
  
\bibitem{Han:2015lha} 
  C.~Han and M.~Park,
  arXiv:1507.07729 [hep-ph].
  
\bibitem{Chalons:2015vja} 
  G.~Chalons and D.~Sengupta,
  JHEP {\bf 1512}, 129 (2015)
  [arXiv:1508.06735 [hep-ph]].
  
  
 \bibitem{Bhattacherjee:2013wna} 
  B.~Bhattacherjee, A.~Choudhury, K.~Ghosh and S.~Poddar,
  ``Compressed supersymmetry at 14 TeV LHC,''
  Phys.\ Rev.\ D {\bf 89}, no. 3, 037702 (2014)
  [arXiv:1308.1526 [hep-ph]].
  
\bibitem{Nath:2016kfp} 
  P.~Nath and A.~B.~Spisak,
  arXiv:1603.04854 [hep-ph].

  
 
  

 
 \bibitem{Giudice:2010wb} 
  G.~F.~Giudice, T.~Han, K.~Wang and L.~T.~Wang,
  Phys.\ Rev.\ D {\bf 81}, 115011 (2010)
  [arXiv:1004.4902 [hep-ph]].
  
\bibitem{Han:2013usa} 
  C.~Han, A.~Kobakhidze, N.~Liu, A.~Saavedra, L.~Wu and J.~M.~Yang,
  JHEP {\bf 1402}, 049 (2014)
  [arXiv:1310.4274 [hep-ph]].
 
 \bibitem{Schwaller:2013baa} 
  P.~Schwaller and J.~Zurita,
  JHEP {\bf 1403}, 060 (2014)
  [arXiv:1312.7350 [hep-ph]].
 
\bibitem{Baer:2014cua} 
  H.~Baer, A.~Mustafayev and X.~Tata,
  Phys.\ Rev.\ D {\bf 89}, no. 5, 055007 (2014)
  [arXiv:1401.1162 [hep-ph]].
 
\bibitem{Han:2014kaa} 
  Z.~Han, G.~D.~Kribs, A.~Martin and A.~Menon,
  Phys.\ Rev.\ D {\bf 89}, no. 7, 075007 (2014)
  [arXiv:1401.1235 [hep-ph]].
  
\bibitem{Bramante:2014dza} 
  J.~Bramante, A.~Delgado, F.~Elahi, A.~Martin and B.~Ostdiek,
  Phys.\ Rev.\ D {\bf 90}, no. 9, 095008 (2014)
  [arXiv:1408.6530 [hep-ph]].
  
  
  \bibitem{Han:2014xoa} 
  C.~Han, L.~Wu, J.~M.~Yang, M.~Zhang and Y.~Zhang,
  Phys.\ Rev.\ D {\bf 91}, 055030 (2015)
  [arXiv:1409.4533 [hep-ph]].
  
  \bibitem{Baer:2014kya} 
  H.~Baer, A.~Mustafayev and X.~Tata,
  Phys.\ Rev.\ D {\bf 90}, no. 11, 115007 (2014)
  [arXiv:1409.7058 [hep-ph]].
  
\bibitem{Bramante:2014tba} 
  J.~Bramante, P.~J.~Fox, A.~Martin, B.~Ostdiek, T.~Plehn, T.~Schell and M.~Takeuchi,
  Phys.\ Rev.\ D {\bf 91}, 054015 (2015)
  [arXiv:1412.4789 [hep-ph]].
  
  \bibitem{Han:2015lma} 
  C.~Han, D.~Kim, S.~Munir and M.~Park,
  JHEP {\bf 1504}, 132 (2015)
  [arXiv:1502.03734 [hep-ph]].
  
\bibitem{Bramante:2015una} 
  J.~Bramante, N.~Desai, P.~Fox, A.~Martin, B.~Ostdiek and T.~Plehn,
  Phys.\ Rev.\ D {\bf 93}, no. 6, 063525 (2016)
  [arXiv:1510.03460 [hep-ph]].
  
  \bibitem{Ismail:2016zby} 
  A.~Ismail, E.~Izaguirre and B.~Shuve,
  arXiv:1605.00658 [hep-ph].
  
\bibitem{Alwall:2014hca} 
  J.~Alwall {\it et al.},
  JHEP {\bf 1407}, 079 (2014)
  [arXiv:1405.0301 [hep-ph]].
  
\bibitem{Ball:2012cx} 
  R.~D.~Ball {\it et al.},
  Nucl.\ Phys.\ B {\bf 867}, 244 (2013)
  [arXiv:1207.1303 [hep-ph]].

\bibitem{Sjostrand:2006za} 
  T.~Sjostrand, S.~Mrenna and P.~Z.~Skands,
  JHEP {\bf 0605}, 026 (2006)
  [hep-ph/0603175].

\bibitem{PGS}
J. Conway et al., \url{http://www.physics.ucdavis.edu/~conway/
research/software/pgs/pgs4‑general.htm}.

\bibitem{Cacciari:2008gp} 
  M.~Cacciari, G.~P.~Salam and G.~Soyez,
  JHEP {\bf 0804}, 063 (2008)
  [arXiv:0802.1189 [hep-ph]].
 
   \bibitem{TrigATLAS8TeV}
  ATL-DAQ-PUB-2012-002
 
  
    \bibitem{TrigCMS13TeV}
\url{https://twiki.cern.ch/twiki/bin/view/CMSPublic/L1TriggerDPGResults}


  
\bibitem{Bramante:2011xd} 
  J.~Bramante, J.~Kumar and B.~Thomas,
  Phys.\ Rev.\ D {\bf 86}, 015014 (2012)
  [arXiv:1109.6014 [hep-ph]].
 
 \bibitem{Gleisberg:2003xi} 
  T.~Gleisberg, S.~Hoeche, F.~Krauss, A.~Schalicke, S.~Schumann and J.~C.~Winter,
  JHEP {\bf 0402}, 056 (2004)
  [hep-ph/0311263].
 
 \bibitem{Gleisberg:2008ta} 
  T.~Gleisberg, S.~Hoeche, F.~Krauss, M.~Schonherr, S.~Schumann, F.~Siegert and J.~Winter,
  JHEP {\bf 0902}, 007 (2009)
  [arXiv:0811.4622 [hep-ph]].
 
  \bibitem{ATLAS13TeVSearch} 
  The ATLAS collaboration,
  ATLAS-CONF-2015-062.

  
  \bibitem{CMS:2015lrd} 
  CMS Collaboration [CMS Collaboration],
  CMS-PAS-SUS-15-002.
  
   \bibitem{CMS1313MT2} 
  CMS Collaboration [CMS Collaboration]
CMS-PAS-SUS-16-015
  
  
  

  
\end{thebibliography}

\end{document}